\documentclass[12pt]{article}
\usepackage{epsfig}

\def\Journal#1#2#3#4{{#1} {\bf #2}, #3 (#4)}

\def\PLB{{\em Phys. Lett.} B}
\def\PRL{\em Phys. Rev. Lett.}
\def\PRD{{\em Phys. Rev.} D}

\def\EPJC{{\em Eur. Phys. J.} C}
\def\PR{\em Phys. Reports}
\def\NPPS{\em Nucl. Phys. Proc. Suppl.}

\def\be{\begin{equation}}
\def\ee{\end{equation}}
\def\bea{\begin{eqnarray}}
\def\eea{\end{eqnarray}}

\begin{document}
%\vspace*{4cm}
\begin{center}
{\Large\bf{Recent positivity constraints for\\
\vskip 0.3cm
 spin observables and parton distributions \footnote{Invited talk presented at the 14th Workshop on Elastic and Diffractive Scattering (EDS Blois Workshop), December 15-21, 2011, Qui Nhon, Vietnam}}}
\vskip1.4cm
{\bf Jacques Soffer}
\vskip 0.2cm
{\it Physics Department, Temple University\\
Barton Hall, 1900 N, 13th Street\\
Philadelphia, PA 19122-6082, USA}
\vskip 0.5cm
\end{center}

\begin{center}
{\bf Abstract}\\
\end{center}
Spin observables allow a deeper understanding of the nature of the underlying dynamics and positivity reduces substantially their allowed domains. We will present some new positivity constraints for spin observables and their implications for parton distributions. We will also make some comparisons with recent data.

\section{Introduction}
Spin observables for any particle reaction, contain some unique information which is very usefull to check the validity of theoretical assumptions. Positivity reduces substantially the allowed domain for spin observables and can be used to determine these domains for exclusive and inclusive spin-dependent reactions.
 We emphasize the relevance of positivity in spin physics, which puts non-trivial model independent constraints on spin
observables. If one, two or several observables are measured, the constraints can help to decide which new observable will provide
the best improvement of knowledge. Different methods can be used to establish these constraints and they have been presented together with many interesting cases in a recent review article \cite{aerst}. Here we will present some new positivity constraints for spin observables and their implications for parton distributions with some comparisons with recent data.\\

\section{The quark transversity distribution $\delta q(x,Q^2)$}
The quark transversity distribution was first mentioned by Ralston and Soper in 1979, in $pp \to \mu^+\mu^- X$, with
transversely polarized protons, but forgotten until 1990, where it was realized that it
completes the description of the quark distribution in a nucleon as a density matrix
\begin{equation}
{\cal Q}(x,Q^2)=q(x,Q^2)I \otimes I + \Delta q(x,Q^2)\sigma_3\otimes\sigma_3
 + \delta q(x,Q^2)(\sigma_+\otimes\sigma_- + \sigma_-\otimes\sigma_+)~.
\label{eq:one}
\end{equation}
So in addition to $q(x,Q^2)$ and $\Delta q(x,Q^2)$, the helicity distribution, there is a new distribution function $\delta q(x,Q^2)$, (also denoted 
$\Delta_{T}q(x,Q^2)$), chiral odd and leading twist,
which decouples from Deep Inelastic Scattering.
By requiring the positivity of the above density matrix, one finds that $\delta q(x,Q^2)$ must satisfy the following positivity bound \cite{js}
\begin{equation}
  \label{eq:two}
  q(x,Q^2) + \Delta q(x,Q^2) \geq 2|\delta q(x,Q^2)| ,
\end{equation}
 which survives up to next-to leading order corrections.\\
\begin{figure}
\begin{center}
%\rule{5cm}{0.2mm}\hfill\rule{5cm}{0.2mm}
%\vskip 2.5cm
\hskip -1.5cm
%\rule{5cm}{0.2mm}\hfill\rule{5cm}{0.2mm}
\psfig{figure=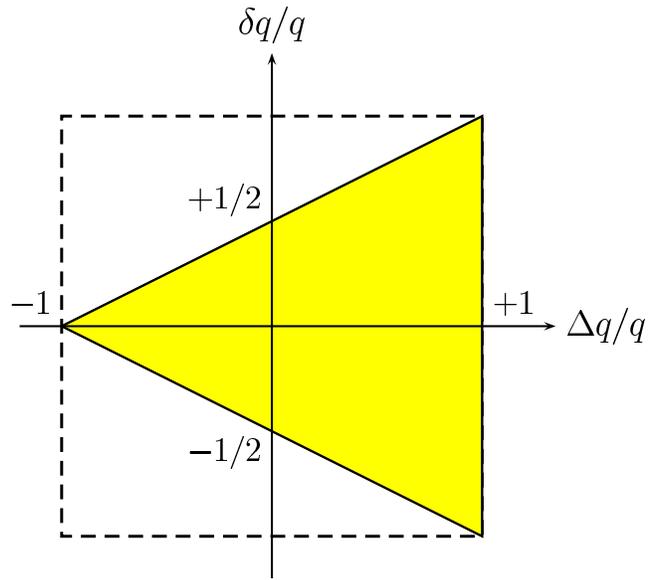,height=2.5in}
\hskip 2.5cm
\psfig{figure=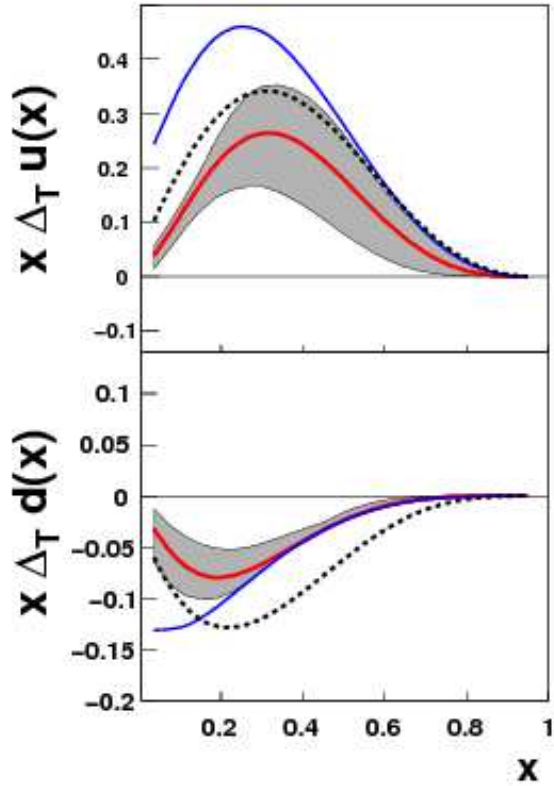,height=5.0in}
\end{center}
\caption{ ({\it Top}): Allowed domain corresponding to the constraint Eq. (\ref{eq:two}). ({\it Bottom}): Transversity distributions for $u$ and $d$ quarks (solid curves inside uncertainties bands), positivity bounds (highest and lowest lines) and helicity distributions (dashed curves) (Taken from Ref. (3)).
\label{fig:js}}
\end{figure}
We show in Fig. \ref{fig:js} (Top), that the region restricted by Eq. (\ref{eq:two}) is only half of the entire square. This theoretical result is not new, but recently it became very interesting to test, with the first extraction of $\delta q$.
The transversity distributions for $u$ and $d$ quarks were determined indirectly from a global analysis \cite{ansel} of the experimental data on azimuthal asymmetries in semiinclusive
deep inelastic scattering (SIDIS), from the HERMES and COMPASS Collaborations, and in $e^{+}e^{-} \to h_{1}h_{2}X$ processes, from the Belle Collaboration. The results at $Q^2 = 2.4\mbox{GeV}^2$ are shown in Fig. \ref{fig:js} (Bottom) and one notices that in both cases $|\Delta_{T}q| \leq |\Delta q|$. It is clear that the uncertainty
on this determination is still rather large. The positivity bound is well respected for the $u$ quark, but saturation, or perhaps violation, seems to occur for
the $d$ quark in the high $x$ region and this is partly related to the fact that $\Delta d$ is large and negative in this region. A special effort should be made on the experimental side to clarify this very important issue.

\section{General positivity bounds in particle inclusive production}

To start, we consider the inclusive reaction of the type  $A(\mbox{spin 1/2})+B(\mbox{spin 1/2})\to C+X$, 
where both initial spin $1/2$ particles can be in any possible directions and no polarization is observed in the final state. The spin-dependent corresponding cross section $\sigma\left(P_a, P_b\right)={\rm Tr}\left(M\rho\right)$, can be defined through the $4\times 4$ cross section matrix $M$ and the spin density matrix $\rho$, 
where $P_a$, $P_b$ are the spin unit vectors of $A$ and $B$, $\rho=\rho_a\otimes \rho_b$ is the spin density matrix with $\rho_a=(I_2+P_a\cdot \vec{\sigma}_a)/2$, and similar for $\rho_b$. Here $I_2$ is the $2\times 2$ unit matrix, and $\vec{\sigma}=(\sigma_x, \sigma_y, \sigma_z)$ stands for the $2\times 2$ Pauli matrices.
For the {\it parity-conserving} process, $M$ could be parametrized in the following way
\bea
M=\sigma_0[I_4+A_{aN}\sigma_{ay}\otimes I_2+A_{bN}I_2\otimes \sigma_{by}
+A_{NN}\sigma_{ay}\otimes \sigma_{by}+A_{LL}\sigma_{az}\otimes \sigma_{bz}\\
\nonumber
+A_{SS}\sigma_{ax}\otimes \sigma_{bx}+A_{LS}\sigma_{az}\otimes\sigma_{bx}
+A_{SL}\sigma_{ax}\otimes\sigma_{bz}].
\label{eq:Mpc}
\eea
Here $I_4$ is the $4\times 4$ unit matrix and $\sigma_0$ stands for the spin-averaged cross
section. In other words, for a parity-conserving process, there are {\it eight} independent
spin-dependent observables: the unpolarized
cross section $\sigma_0$, {\it two} single transverse spin asymmetries $A_{aN}$ and $A_{bN}$, and {\it five} double spin asymmetries $A_{NN}$, $A_{LL}$, $A_{SS}$, $A_{LS}$, and $A_{SL}$. 
Here the subscript $L$, $N$, $S$ represents the unit vectors along the spin directions of initial particles $A$ and $B$. Specifically in the center-of-mass system of $A$ and $B$, $L$, $N$, $S$ are along the incident momentum, along the normal to the scattering plane which contains $A$, $B$ and $C$, and along $N\times L$, respectively.\\
The crucial point is that $M$ is a Hermitian and positive matrix and in order to 
derive the positivity conditions. In the transverse basis where $\sigma_y$ is diagonal, we have found that the diagonal matrix elements $M_{ii}$ are given by

\bea
M_{11}=(1+A_{NN})+(A_{aN}+A_{bN}),~ M_{22}&=&(1-A_{NN})+(A_{aN}-A_{bN})\\
\nonumber
%\left
M_{33}=(1-A_{NN})-(A_{aN}-A_{bN}),~ M_{44}&=&(1+A_{NN})-(A_{aN}+A_{bN}).
\label{eq:pos1}
\eea
Since one of the necessary conditions for a Hermitian matrix to be positive definite is that all the diagonal matrix elements has to be positive $M_{ii}\geq 0$, we thus derive
$1\pm A_{NN}\geq \left|A_{aN}\pm A_{bN}\right|$, which is valid in full generality, for both parity-conserving and parity-violating processes.\\
Back to the case for $p^\uparrow+p^\uparrow\to C+X$ where the initial particles are identical, we have $A_{aN}(y)=-A_{bN}(-y)$. Using this relation , one obtains,
\be
\
1 \pm A_{NN} (y) \geq \left|A_N (y)\mp A_N(-y)\right|.
\label{pos2}
\ee

\begin{figure}
\begin{center}
%\rule{5cm}{0.2mm}\hfill\rule{5cm}{0.2mm}
%\vskip 2.5cm
%\rule{5cm}{0.2mm}\hfill\rule{5cm}{0.2mm}
\hskip 5mm
\psfig{figure=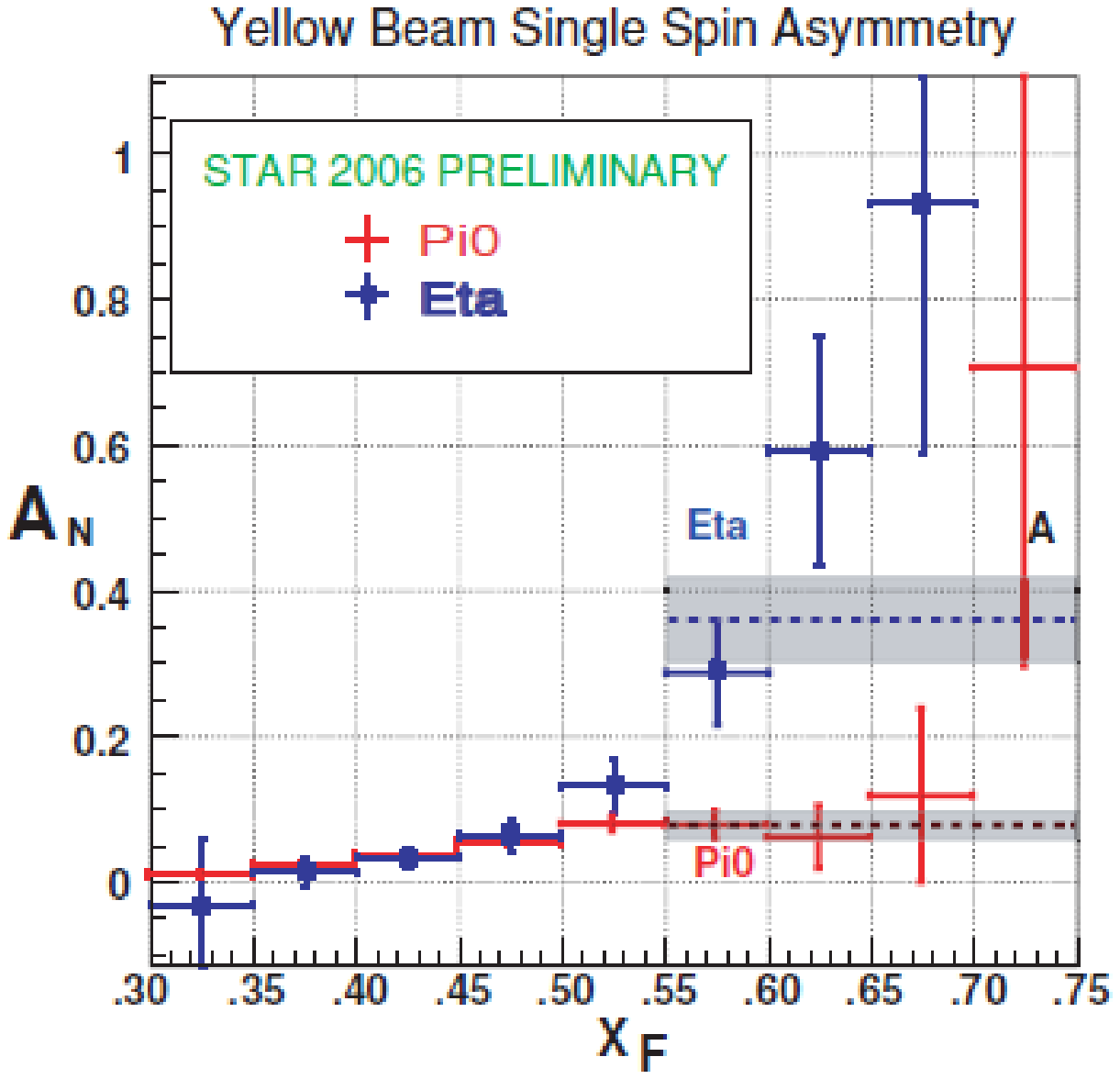,height=3.2in}
\hskip 15mm
\psfig{figure=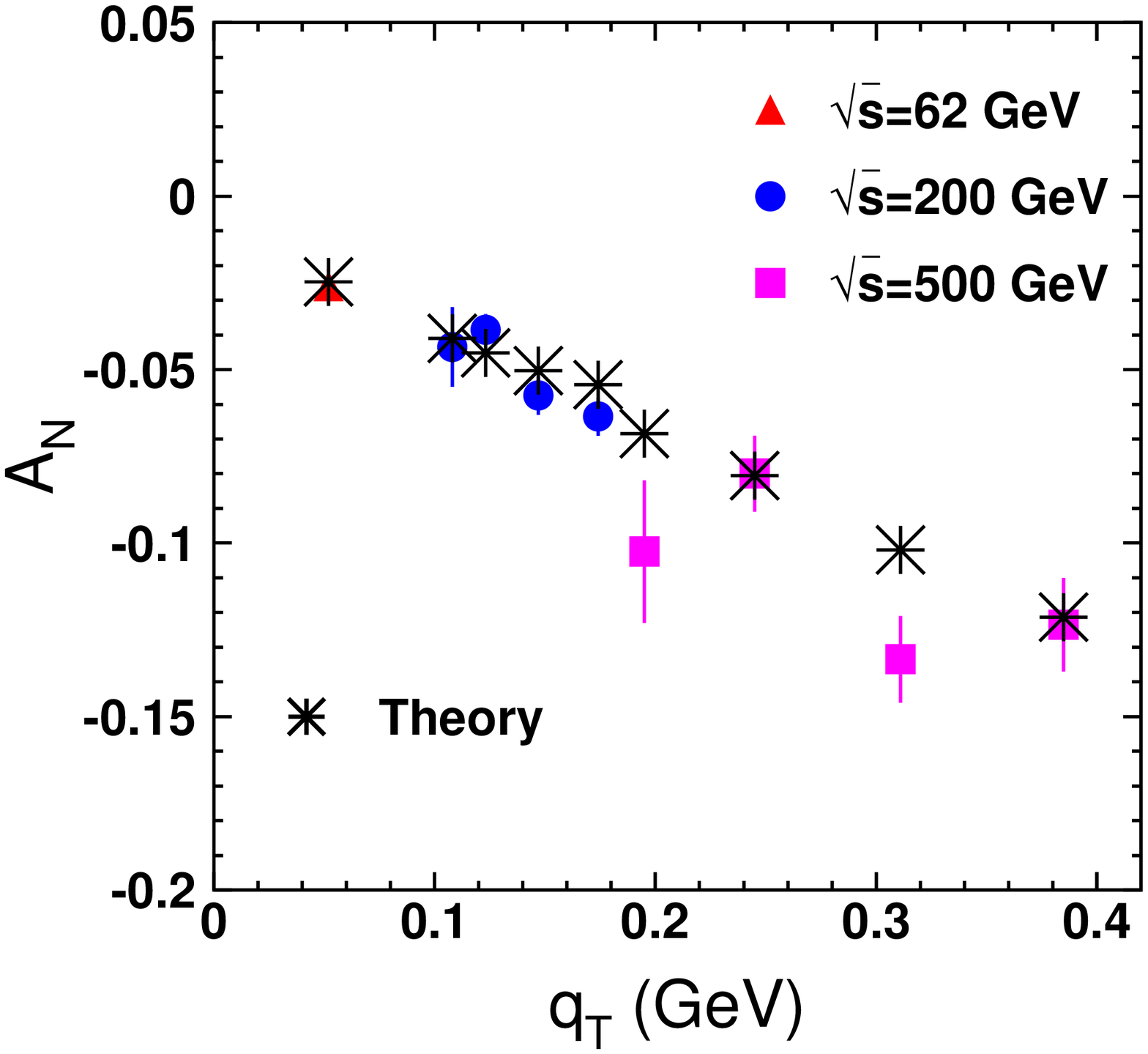,height=2.8in}
\end{center}
\caption{({\it Top}): The dependence of $A_N$ on $x_F$ is plotted for $2\gamma$ production in the $\pi^0$ and $\eta$ mass regions (Taken from Ref. [5]). ({\it Bottom}): $A_N$ for the reaction $pp \to nX$ versus the neutron transverse momentum $q_T$, measured at various energies. The asterisks are the result of the calculation (Taken from Ref. [6]).
\label{fig:An}}
\end{figure}
This is an interesting result which, can be used, in principle, with available data on $A_N$ for $\pi^{\pm}$, $K^{\pm}$, $\pi^0$, $\eta$ production, to put some non trivial contraints on $A_{NN}(y)$. For illustration we display in Fig. 2 some recent unexpected and large results on single spin asymmetries. In the case of neutron production, the asymmetry corresponds to very forward
neutrons and the big effect observed is due to an interference of $\pi -a_1$ Regge exchanges \cite{kpss}.\\
Conversely, for prompt photon and jet production, theoretically $A_{NN}$ is expected to be small, except in the high $p_T$ region, because since there is no transversity for the gluons, its sensitivity to $\delta q$ is only in the high $p_T$ region, dominated by
 $qq \to qq$. By making use of the positivity bound on transversity, it was possible to estimate $A_{NN}$ \cite{ssv} and it was observed that $|A_{NN}|<<|A_{LL}|$. It is important to verify this theoretical expectation.
\section{Positivity bounds for Sivers functions}
Now let us study the implication of $1 \pm A_{NN} (y) \geq \left|A_N (y)\pm A_N(-y)\right|$, in the parity-violating process $p^\uparrow+p^\uparrow\to W^{\pm}+X$. We denote by $y$ and $\mathbf{q}_{\perp}$ the rapidity and transverse momentum of the $W$ boson. Since we will assume $A_{NN}\approx 0$, for $y=0$, the bound reduces to
$1 \geq 2|A_N (y = 0)|$, to be compared with the usual trivial bound $1 \geq |A_N (y = 0)|$.\\
The TMD quark distribution in a transversely polarized hadron of spin $\vec{S}$, can be expanded as
\be
\label{sivers}
f_{q/h^\uparrow}(x,\mathbf{k}_{\perp},\vec{S})
\equiv 
f_{q/h}(x,k_{\perp}) +  
\frac{1}{2}\Delta^N f_{q/h^\uparrow}(x,k_\perp)\,
\vec{S}\cdot \left(\hat{p}\times \hat{\mathbf{k}}_\perp \right)\,,
\ee
where $\hat{p}$ and $\hat{\mathbf{k}}_\perp$ are the unit vectors of $\vec{p}$ and 
$\mathbf{k}_\perp$, respectively. $f_{q/h}(x,k_\perp)$ is the spin-averaged TMD distribution, and $\Delta^N f_{q/h^\uparrow}(x,k_\perp)$ is the Sivers function. There is a trivial positivity bound for the Sivers functions which reads 
$|\Delta^N f_{q/h^\uparrow}(x,k_\perp)|\leq 2f_{q/h}(x,k_{\perp}$). $A_N$ is expressed by an integral in terms of the Sivers functions, and it is usually assumed in the phenomenological studies that the $x$ and $k_\perp$ dependence of the TMD distributions can be further factorized as follows 
\bea
\label{fact}
f_{q/h}(x,k_\perp) 
&=& f_q(x)g(k_\perp)\,,
\\
\nonumber
\Delta^N f_{q/h^\uparrow}(x,k_\perp)
&=&
\Delta^N f_{q/h^\uparrow}(x)\,h(k_\perp)g(k_\perp).
\eea
For the $k_\perp$ dependence, a Gaussian ansatz is 
usually introduced , which has the form
\bea
\label{gauss}
g(k_\perp)&=&\frac{1}{\pi\langle k_\perp^2\rangle}
e^{-k_\perp^2/\langle k_\perp^2\rangle}~,
\\
\nonumber
h(k_\perp)
&=&
\sqrt{2e}\, \frac{k_\perp}{M_1}
e^{-k_\perp^2/M_1^2}~.
\eea
This choice satisfies $h(k_\perp)\leq 1$. 
Thus the previous positivity bound implies
$\left|\Delta^N f_{q/h^\uparrow}(x)\right|\leq  2f_q(x)$. 
Then one can carry out the integration analytically in $A_N$ to obtain
\be
A_N(y=0) = H(q_\perp)
\frac{\sum_{ab}|V_{ab}|^2 
      \Delta^N f_{a/p^\uparrow}(x)\,f_{b}(x)}
     {\sum_{ab}|V_{ab}|^2\,f_{a}(x)\, f_{b}(x)} ,
\label{An_W}
\ee
where $x=M_W/\sqrt{s}$ for $y=0$ and $H(q_\perp)$ is given by
\be
\label{H}
H(q_\perp)=\vec{S}_\perp \cdot (\hat{p}\times \mathbf{q}_\perp)\,
\frac{\sqrt{2e}}{M_1}
\frac{\langle k_s^2\rangle^2}
     {[\langle k_\perp^2\rangle + \langle k_s^2\rangle]^2}\,
     e^{-\left[ \frac{\langle k_\perp^2\rangle-\langle k_s^2\rangle}
     {\langle k_\perp^2\rangle + \langle k_s^2\rangle}\right] 
     \frac{\mathbf{q}_\perp^2}{2\langle k_\perp^2\rangle}},
\ee
with $\langle k_s^2\rangle=M_1^2\, \langle k_\perp^2\rangle
/[M_1^2+\langle k_\perp^2\rangle]$.\\
The $q_\perp$-dependent function $H(q_\perp)$ reaches its 
maximum $H(q_\perp)_{\rm max}$ when 
$q_\perp^2=\langle k_\perp^2\rangle(\langle k_\perp^2\rangle + \langle k_s^2\rangle)
/(\langle k_\perp^2\rangle - \langle k_s^2\rangle)$, 
with $H(q_\perp)_{\rm max}$ given by
\be
\label{Hmax}
H(q_\perp)_{\rm max}=\frac{\langle k_s^2\rangle^2}
     {[\langle k_\perp^2\rangle + \langle k_s^2\rangle]^2}
     \left[\frac{2\langle k_\perp^2\rangle}{M_1^2}
     \frac{\langle k_\perp^2\rangle + \langle k_s^2\rangle}
     {\langle k_\perp^2\rangle - \langle k_s^2\rangle}\right]^{\frac{1}{2}}.
\ee
Using the fact that $1\geq 2|A_N (y = 0)|$ for any $q_\perp$ and 
$\sqrt{s}$, we thus derive a new bound for the Sivers functions
\be
\frac{\left|\sum_{ab}|V_{ab}|^2 
      \Delta^N f_{a/A^\uparrow}(x)\,f_{b}(x)\right|}
     {\sum_{ab}|V_{ab}|^2\,f_{a}(x)\, f_{b}(x)}\leq 
     \frac{1/2}{H(q_\perp)_{\rm max}}.
\label{newbound}     
\ee
For $W^+$, it can be simplified as $\left|\frac{\Delta^N u(x)}{u(x)}+\frac{\Delta^N \bar{d}(x)}{\bar{d}(x)}\right|
\leq \frac{1}{H(q_\perp)_{\rm max}}$. For $W^-$, one obtains the following constraint $\left|\frac{\Delta^N d(x)}{d(x)}+\frac{\Delta^N \bar{u}(x)}{\bar{u}(x)}\right|\leq \frac{1}{H(q_\perp)_{\rm max}}$. For $\langle k_\perp^2\rangle=0.25~{\rm GeV}^2$ and $M_1^2=0.34^{+0.30}_{-0.16}~{\rm GeV}^2$, one gets $2.6<1/H(q_\perp)_{\rm max}<4.75$, to be compared with the trivial bound, which gives the number 4 on the r.h.s. Although it is not a spectacular result, these new bounds are useful for consistency checks and we note that
this can be also applied to the Sivers gluon function accessible in direct photon production \cite{ssy}.

\section{Positivity bounds involving parity-violating asymmetries}
\begin{figure}
\begin{center}
%\rule{5cm}{0.2mm}\hfill\rule{5cm}{0.2mm}
%\vskip 2.5
%\rule{5cm}{0.2mm}\hfill\rule{5cm}{0.2mm}
\hskip -5mm
\psfig{figure=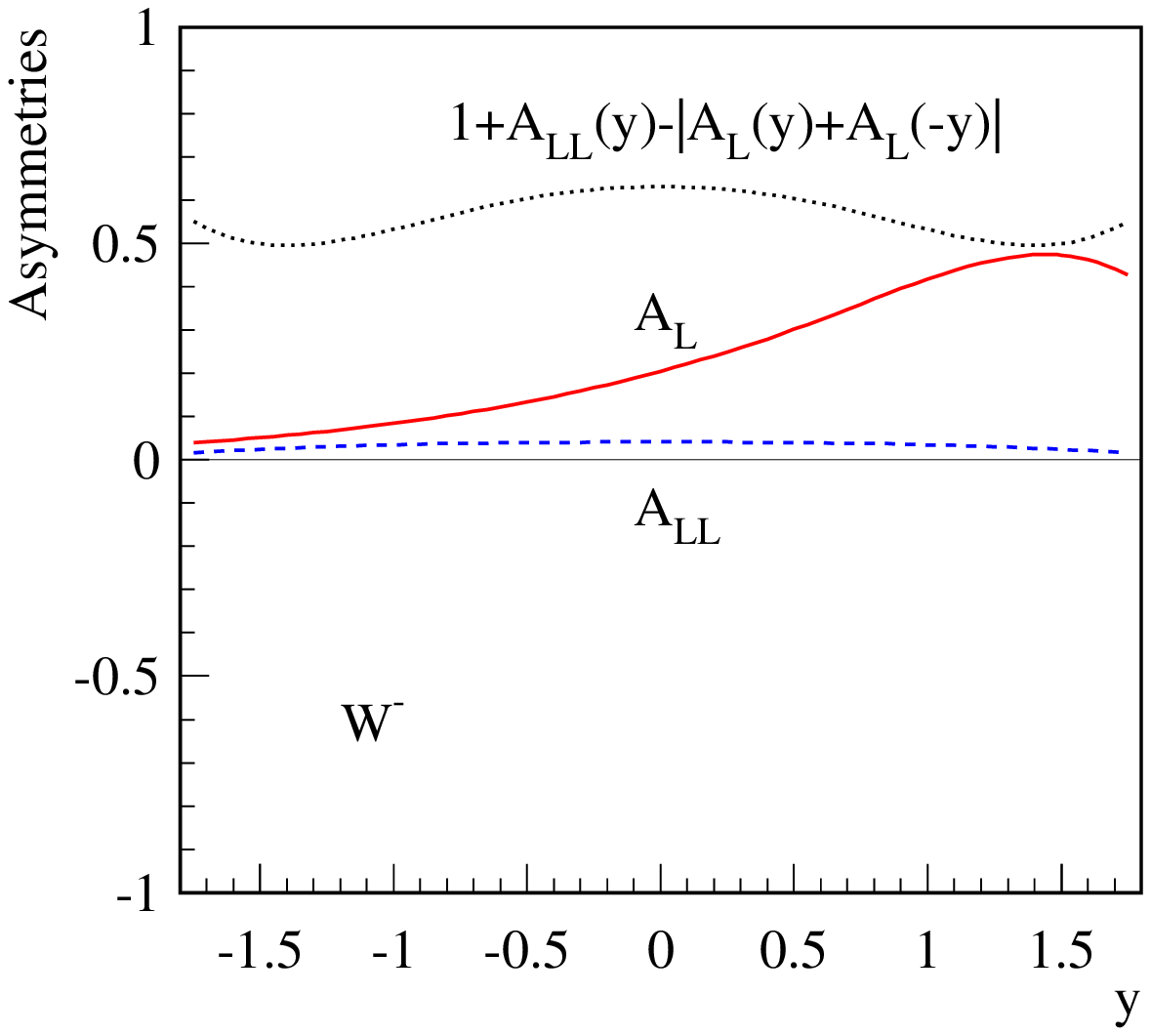,height=2.4in}
\hskip 0.5cm
\psfig{figure=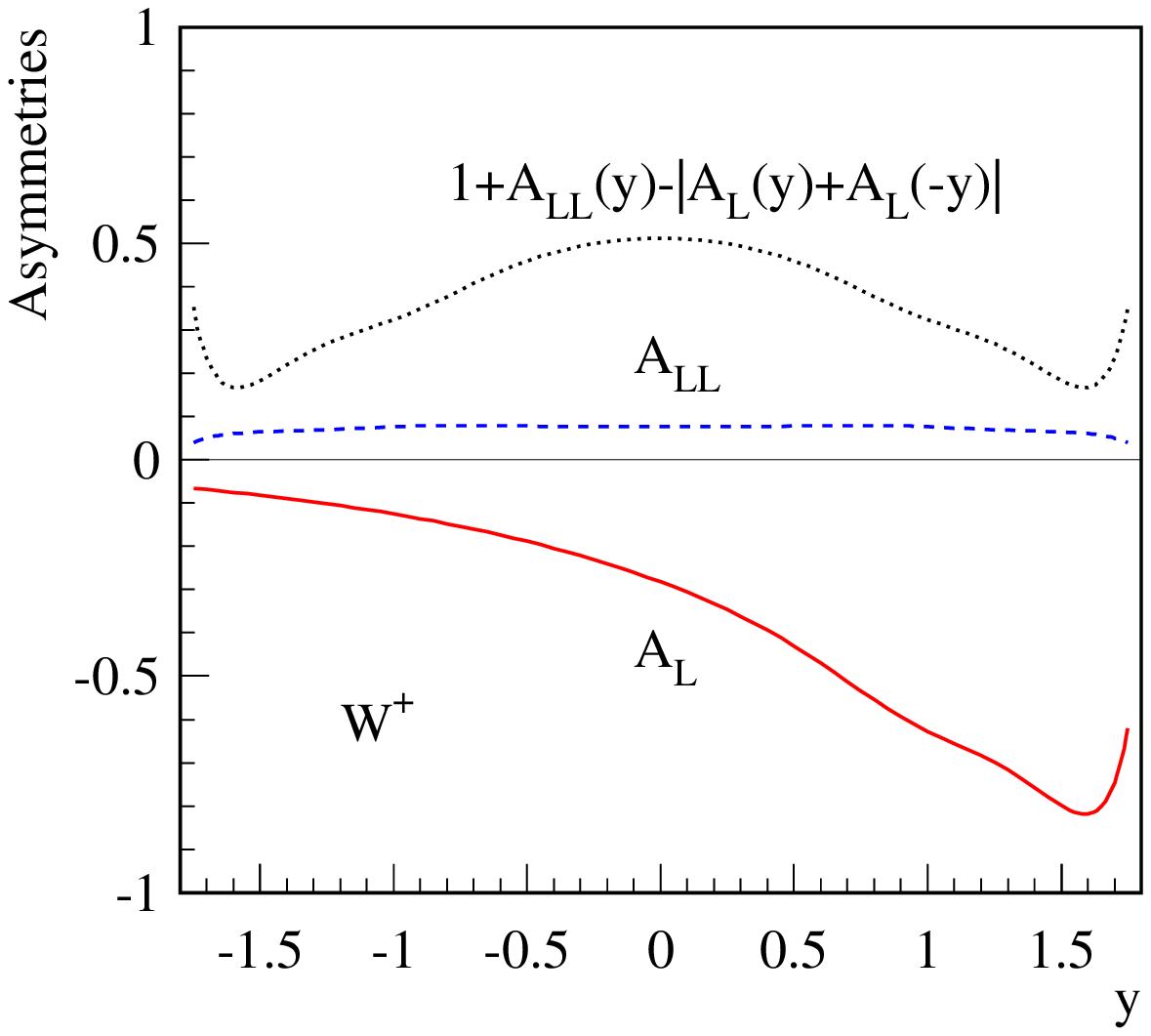,height=2.4in}
\hskip 0.5cm
\psfig{figure=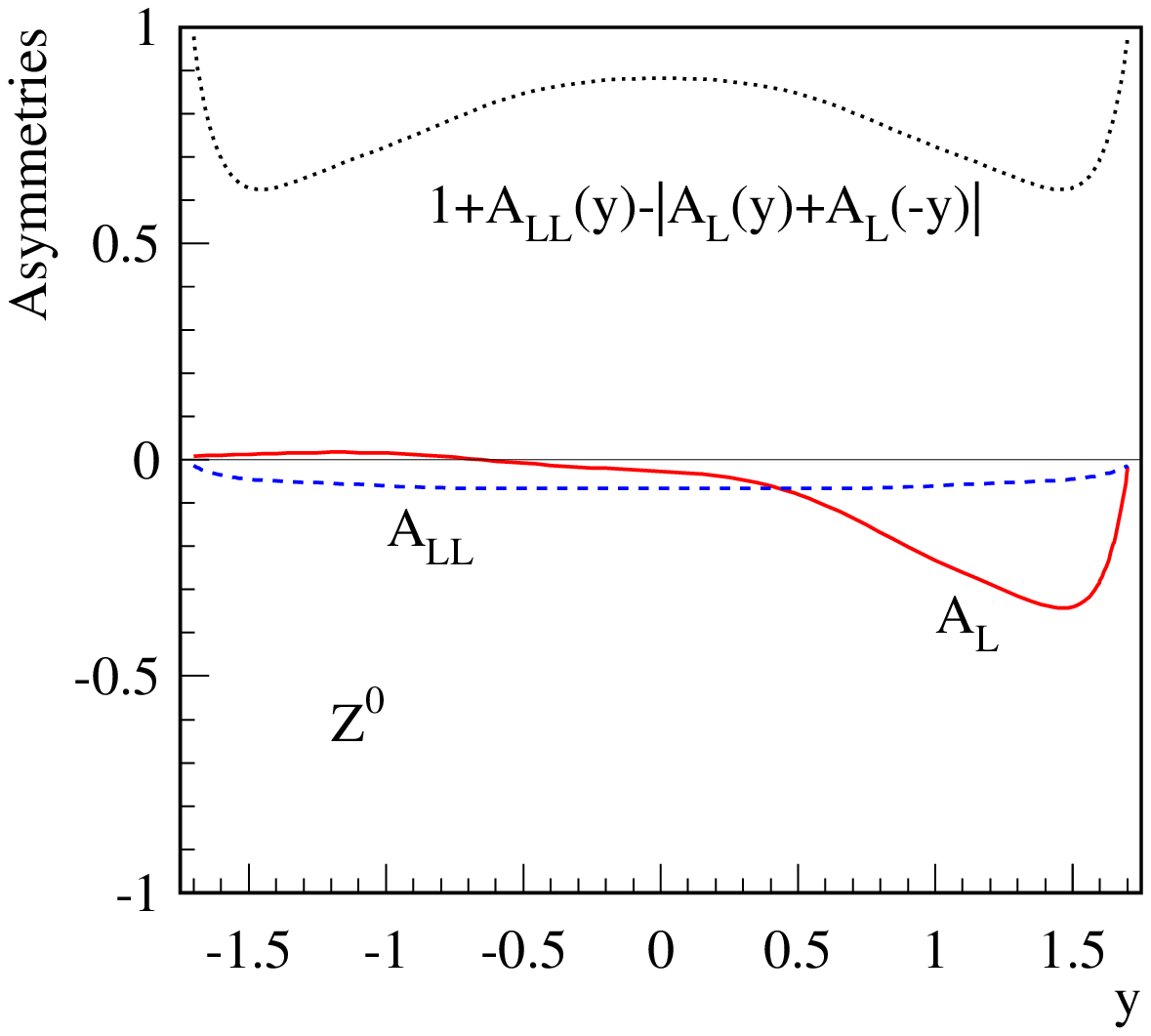,height=2.4in}
\end{center}
\caption{Longitudinal asymmetries are plotted as a function of rapidity $y$ of the $W^-$ ({\it Top Left}), $W^+$ ({\it Top Right}), $Z^0$ ({\it Bottom}), produced in $pp$ collisions at $\sqrt{s}=500 \mbox{GeV}$. The solid curves are the single longitudinal spin asymmetry $A_L$, the dashed curves are the double longitudinal spin asymmetry $A_{LL}$ and the dotted curves are the combination of $1 + A_{LL}(y)-|A_{L}(y) + A_{L}(-y)|$ (Taken from Ref. [4]).
\label{fig:An}}
\end{figure}

In the helicity basis it is easy to obtain the explicit form of $M$ and now from $M_{ii}\geq 0$, we have
$1\pm A_{LL}>|A_{aL}\pm A_{bL}|$. It is important to note that for identical initial particles scattering, one has
$A_{aL}(y)=A_{bL}(-y)$, so one gets
\be
\label{pv}
1\pm A_{LL}(y)>|A_{L}(y)\pm A_{L}(-y)|.
\ee
These bounds should be tested in RHIC experiments for $W^{\pm}$ or $Z^0$ production in longitudinal $pp$ collisions, $\vec{p}\vec{p}\to W^{\pm}/Z^0+X$. In perturbative QCD formalism, at leading-order and restricting to only up and down quarks, one has the following simple expressions for the single helicity asymmetries
\bea
\label{pv1}
A^{W^+}_{L} (y)& = & \frac{-\Delta u(x_a) \bar d(x_b) +
\Delta \bar d(x_a) u(x_b)}{u(x_a) 
\bar d(x_b) + \bar d(x_a) u(x_b)}\; , 
\\
\label{pv2}
A^{W^-}_{L} (y)& = & \frac{- \Delta d(x_a) \bar u(x_b) +
\Delta \bar u(x_a) d(x_b)}{d(x_a) 
\bar u(x_b) + \bar u(x_a) d(x_b)}\; ,  
\\
\label{pv3}
A^{Z^{0}}_{L} (y) & = & \frac{\sum_q (2 v_q a_q) 
\left[-\Delta q(x_a) \bar q(x_b) +
\Delta \bar q(x_a) q(x_b) \right] }
{ \sum_q (v_q^2 + a_q^2) 
\left[  q(x_a) \bar q(x_b) + 
\bar q(x_a)  q(x_b) \right] }\; ,
\eea
and for the double helicity asymmetries
\bea
\label{all1}
A^{W^+}_{LL} (y)& = & - \frac{\Delta u(x_a) \Delta \bar d(x_b) +
\Delta \bar d(x_a) \Delta u(x_b)}{u(x_a) 
\bar d(x_b) + \bar d(x_a) u(x_b)}\; , 
\\
\label{all2}
A^{W^-}_{LL} (y)& = & - \frac{\Delta d(x_a) \Delta \bar u(x_b) +
\Delta \bar u(x_a) \Delta d(x_b)}{d(x_a) 
\bar u(x_b) + \bar u(x_a) d(x_b)}\; ,
\\
\label{all3}
A^{Z^{0}}_{LL} (y) & = & - \frac{\sum_q (v_q^2 + a_q^2) 
\left[ \Delta q(x_a) \Delta \bar q(x_b) + 
\Delta \bar q(x_a) \Delta q(x_b) \right] }
{ \sum_q (v_q^2 + a_q^2) 
\left[  q(x_a) \bar q(x_b) + 
\bar q(x_a)  q(x_b) \right] }\; ,
\eea
where $\Delta q(x)$ and $q(x)$ are the helicity distribution and unpolarized parton distribution function, respectively. $v_q$ and $a_q$ are the vector and axial couplings of the $Z^0$ boson to the quark. $x_{a,b}$ are the parton momentum fractions given by
$x_a=m_Q/\sqrt{s} \,e^{y}$, $x_b=m_Q/\sqrt{s} \,e^{-y}$,
with $m_Q$ and $y$, the mass and rapidity of the $W$ (or $Z$) boson and $\sqrt{s}$ the center-of-mass energy.\\
One notices that $A_L$ and $A_{LL}$ involve only quark helicity distributions. To estimate these asymmetries numerically, we choose the unpolarized and polarized quark ditributions functions based on a statistical approach \cite{bbs}. The results of the calculations are displayed in Fig. 3 and show that positivity is satisfied. It is not always the case for other sets of parton distributions \cite{zkjs}.

\section{Concluding remarks}
Although many interesting cases were not presented here, for lake of time, we have seen on these recent results that positivity provides very useful non-trivial bounds on spin observables and parton distributions. One should not forget this important tool to check
the consistency of model builders and of forthcoming data, in the exciting field of spin physics.

\section*{Acknowledgments}
I am grateful to Prof. Chung-I Tan for organizing such an interesting scientific program. My warmest congratulations go to Prof. Tran Thanh Van for his wonderful project, in Quy Nhon, of the International Center for Interdisciplinary Science and Education (ICISE) in Vietnam, which will become soon a reality.

%\section*{References}

\end{document}